\documentclass[epj]{svjour}   

\usepackage{times}
\usepackage{amsmath}

\input epsf

\numberwithin{equation}{section} 
\newcounter{append}
\newcommand\Appendix[1]{
 \setcounter{equation}{0}
 \setcounter{subsection}{0}
 \addtocounter{append}{1}
 \def\theappend{\Alph{append}}
 \def\thesection{\theappend}
 \renewcommand{\theequation}{\theappend.\arabic{equation}}
 \section*{Appendix \theappend: #1}}
\begin{document}

\title{Aperiodic extended surface perturbations in the Ising model}

\author{L. Turban\thanks{\email{turban@lps.u-nancy.fr}}}                    

\institute{Laboratoire de Physique des Mat\'eriaux\thanks{The 
Laboratoire de Physique des Mat\'eriaux is the Unit\'e Mixte 
de Recherche C.N.R.S. N$^\circ$7556.}, Universit\'e  Henri Poincar\'e, BP
239, 54506 Vand\oe uvre l\`es Nancy cedex,
France}                                

\date{Received: date / Revised version: date} 

\abstract{We study the influence of an aperiodic extended surface 
perturbation on the surface critical behaviour of the two-dimensional 
Ising model in the extreme anisotropic limit. The perturbation decays 
as a power $\kappa$ of the distance $l$ from the free 
surface with an oscillating amplitude $A_l\!=\!(-1)^{f_l}$ where 
$f_l\!=\!0,1$ follows some aperiodic sequence with an asymptotic 
density equal to $1/2$ so that the mean ampltitude vanishes. 
The relevance of the perturbation is discussed by combining scaling 
arguments of Cordery and Burkhardt for the Hilhorst-van Leeuwen model 
and Luck for aperiodic perturbations. The relevance-irrelevance 
criterion involves the decay exponent $\kappa$, 
the wandering exponent $\omega$ which governs the fluctuation of the 
sequence and the bulk correlation length exponent $\nu$. Analytical
results are obtained for the surface magnetization which displays a 
rich variety of critical behaviours in the $(\kappa,\omega)$-plane.
The results are checked through a numerical finite-size-scaling study. 
They show that second-order effects must be taken into account
in the discussion of the relevance-irrelevance criterion. The scaling 
behaviours of the first gap and the surface energy are also discussed.
} 

\maketitle

\section{Introduction}

\label{intro}
At the free surface of a homogeneous $d$-dimensional sytem with short-range
interactions displaying a bulk second-order phase transition, the scaling
dimension $x_{\rm e}^{\rm s}$ of the surface energy density is equal to
$d$~\cite{burkhardt87}. As a consequence, a weak short-range surface perturbation
$\Delta K_{\rm s}$ of the reduced interaction $K\!=\!\beta J\!=\!J/k_{\rm B}T$
cannot change the surface critical behaviour, since its scaling
dimension $y_{\Delta K}\!=\!d-1-x_{\rm e}^{\rm s}\!=\!-1$ and the
perturbation is irrelevant. Such a pertubative argument
does not exclude the occurence in $d\!>\!2$ dimensions of special,
extraordinary, and surface transitions for strong enough
enhancement of the surface couplings~\cite{binder83}. 

The situation is somewhat different for the Hilhorst-van Leeuwen
(HvL) model~\cite{hilhorst81}, in which the surface perturbation extends 
into the bulk of the system, decaying as a power of the distance $l$
from the surface with:  
\begin{equation}  
\Delta K(l)=\frac{A}{l^\kappa}\,.
\label{deltaj}
\end{equation}  
Such extended surface perturbations may change the surface
critical behaviour for an arbitrarily small 
value of the perturbation amplitude $A$. This can be 
explained~\cite{cordery82} by
noticing that, under a length scale transformation $l'\!=\!l/b$, the
extended thermal perturbation in~(\ref{deltaj}) transforms as:
\begin{equation}  
[\Delta K(l)]'=\frac{A'}{{l'}^\kappa}=b^{1/\nu}\frac{A}{l^\kappa}\,,
\label{scaldelta}
\end{equation}  
so that, comparing both sides of the last equation, the
perturbation amplitude scales as:
\begin{equation}  
A'=b^{-\kappa+1/\nu}\, A\,.
\label{scala}
\end{equation}  
When the decay is strong enough ($\kappa\!>\!1/\nu$), the perturbation
is irrelevant: it does not affect the surface critical behaviour.
When $\kappa\!<\!1/\nu$, the perturbation is relevant: the decay is
sufficiently slow to modify the surface critical behaviour. In the
marginal situation, $\kappa\!=\!1/\nu$, one expects a nonuniversal
surface critical behaviour. 

Analytical results obtained for the
two-dimensional ($2d$) Ising model, with $\nu\!=\!1$, are in complete agreement
with these predictions~\cite{blote83,hvl}: 

i)  For $\kappa\!>\!1$, the scaling
dimensions of the surface magnetization and the surface energy
density are $x_{\rm m}^{\rm s}\!=\!1/2$ and $x_{\rm e}^{\rm s}\!=\!2$,
respectively, like in the homogeneous semi-infinite system. 

ii) In the marginal case, $\kappa\!=\!1$, when $A$ is smaller than a
critical value $A_{\rm c}$, the surface transition is second-order
with continuously varying surface scaling dimensions $x_{\rm
m}^{\rm s}(A)$ and $x_{\rm e}^{\rm s}(A)$. For $A\!>\!A_{\rm c}$, the
transition is first-order: there is a nonvanishing spontaneous
surface magnetization at the bulk critical point which vanishes
above $T_{\rm c}$ since the surface is one-dimensional.

iii) For $\kappa\!<\!1$, the surface transition is first-order for
$A\!>\!0$ and continuous for $A\!<\!0$ where the surface magnetization
displays an essential singularity. 

In the present work we study an aperiodic version of the HvL
model where the amplitude of the decay is modulated according to
some aperiodic sequence. More specifically, we consider a
layered  semi-infinite system with constant reduced interactions
$K_1\!=\!\beta J_1$ between nearest neighbours in the directions
parallel to the free surface while the reduced interactions 
in the perpendicular direction are modulated and reads:
\begin{equation}  
K_2(l)=K_2+\frac{A_l}{l^\kappa}\,,\qquad A_l=(-1)^{f_l}\, A\,,
\label{k2l}
\end{equation}  
where $f_l$ is aperiodic and takes the values 0 or 1. The aperiodic
sequence is assumed to have equal densities of 0 and 1, such that
the mean amplitude vanishes. Otherwise, the critical behaviour would be
the same as for the HvL model with $A$ replaced by the average of
the modulation amplitude $A_l$.

A relevance-irrelevance criterion will be proposed, which is a
combination of the arguments of Burkhardt and Cordery for HvL
perturbations~\cite{cordery82} and those of Luck for aperiodic
systems~\cite{luck93}.

In Section~\ref{sec:1}, we recall the main results about
the fluctuation properties of aperiodic sequences generated via
substitutions and derive a relevance-irrelevance criterion for
aperiodic HvL perturbations, which is valid to linear order in the
perturbation amplitude. In Section~\ref{sec:2}, we examine the
critical behaviour of the surface magnetization in the aperiodic
HvL Ising quantum chain. The results of
Section~\ref{sec:2} are confronted to numerical data obtained via
finite-size scaling (FSS) for different aperiodic sequences in
Section~\ref{sec:3}. The influence of higher-order terms on the 
relevance of the perturbation, the scaling behaviours of the first 
gap and the surface energy are discussed in Section~\ref{sec:4}.
Technical details are given in Appendices A and B.

\section{Fluctuation of the perturbation amplitude and
relevance-irrelevance criterion} 

\label{sec:1}

We consider the aperiodic HvL model defined in~(\ref{k2l}) for a
$d$-dimensional semi-infinite system with a bulk correlation length
exponent $\nu$. The modulation of the perturbation amplitude 
follows some aperiodic sequence which can be generated through
substitution rules~\cite{queffelec87}. 

Working with a finite alphabet $A$, $B$, $C$, ... and starting for
example with $A$, an infinite sequence of letters is obtained by
first substituting  a finite word $S(A)$ for $A$ and then iterating
the process (with $B\to S(B)$, $C\to S(C)$, ...). The
required aperiodic sequence of digits $f_l$ is finally obtained by
replacing each of the letters by a corresponding finite
sequence of digits 0 and 1. 

The information about the properties of the sequence is contained
into the substition matrix with entries $n_{ij}$ giving the numbers
of letters $i$ in $S(j)$ ($i$, $j\!=\!A$, $B$, $C$, ...). Let
$V_\alpha$ be the right eigenvectors of the substitution matrix and
$\Lambda_\alpha$ the corresponding eigenvalues. The leading
eigenvalue, $\Lambda_1$, is related to the dilatation factor $b$ of the
self-similar sequence and gives its length
after $n$ substitutions through $L_n\sim\Lambda_1^n$. 

The asymptotic densities of letters $i\!=\!A$, $B$, $C$, ... are related
to the components of the eigenvector corresponding to the leading
eigenvalue through: 
\begin{equation}  
\rho_\infty(i)=\frac{V_1(i)}{\sum_j V_1(j)}\,,
\label{rho}
\end{equation}  
which allows a calculation of $\rho_\infty$, the asymptotic density
of 1 in the sequence of digits $f_l$.

The sum of the digits $n_l\!=\!\sum_{k=1}^l f_k$ is such that:
\begin{equation}  
n_l-l\rho_\infty\sim l^\omega\,,
\label{nl1}
\end{equation}  
with an amplitude which is log-periodic, i.e.,  a periodic function 
of $\ln l/\ln b$. The exponent
\begin{equation}  
\omega=\frac{\ln\vert\Lambda_2\vert}{\ln\Lambda_1}
\label{omega}
\end{equation}  
is the wandering exponent of the sequence. When $\omega\!<\!0$($>\!0$) the
sequence has  bounded (unbounded) fluctuations.

Using (\ref{nl1}) and the identity
\begin{equation}
(-1)^{f_k}=1-2f_k\,, 
\label{ident}
\end{equation}
the amplitude of the aperiodic HvL perturbation in~(\ref{k2l}),
ave\-ra\-ged at a length scale $l$, can be written as:
\begin{equation}
\overline{A}(l)=\frac{A}{l}\sum_{k=1}^l (-1)^{f_k}
=A(1-2\frac{n_l}{l})\sim Al^{\omega-1}\,,
\label{aav}
\end{equation}
when the asymptotic density of 1 is $\rho_\infty\!=\!1/2$.

To analyse the relevance of the perturbation, one may
replace $A_l$ by $Al^{\omega-1}$ in~(\ref{k2l}), which leads to the
following behaviour under rescaling:
\begin{equation}  
[\Delta K_2(l)]'=\frac{A'}{{l'}^{\kappa-\omega+1}}
=b^{1/\nu}\frac{A}{l^{\kappa-\omega+1}}\,,
\label{scaldelta2}
\end{equation}  
or, for the perturbation amplitude:
\begin{equation}  
A'=b^{-\kappa+\omega-1+1/\nu}A\,.
\label{scala1}
\end{equation}  

Thus, to the first order in the amplitude, the
perturbation is relevant (irrelevant) when 
$\kappa\!<\!(>)\omega-1+1/\nu$. Nonuniversal
surface critical behaviour is expected in the marginal
situation where $\kappa\!=\!\omega-1+1/\nu$.

 For the Ising model in $2d$, $\nu\!=\!1$, and the relevance of the
aperiodic HvL perturbation is simply governed by the sign of
$\omega-\kappa$.

One may notice that Luck's criterion for aperiodic 
perturbations~\cite{luck93}
is recovered when $\kappa\!=\!0$, which corresponds to an
aperiodic modulation with a constant amplitude. 
  
\section{Surface magnetization of the Ising quantum
chain: analytical approach}

\label{sec:2}

In this section, we study the critical behaviour of the
surface magnetization for the $2d$ Ising model with an 
aperiodic extended surface perturbation given by~(\ref{k2l}).

\subsection{Ising quantum chain in a transverse field} 
\label{sec:2.1}

We work in the extreme
anisotropic limit~\cite{kogut79} where $K_1\to\infty$, $K_2(l)\to0$,
$A\to0$, while the ratios 
\begin{equation}  
\lambda=K_2/K_1^*\,,\qquad a=A/K_2\,, 
\label{para}
\end{equation}  
are kept fixed. The dual interaction $K_1^*\!=\!-1/2\ln(\tanh K_1)$ 
enters into the expression of the unperturbed coupling $\lambda$.

In the extreme anisotropic limit, the row-to-row transfer 
operator, ${\cal T}\!=\!\exp(-2K_1^*{\cal H})$, involves  the
Hamiltonian of the quantum Ising chain in a transverse
field~\cite{kogut79,pfeuty70}:    
\begin{equation}  
{\cal H}=
-\frac{1}{2}\sum_{l=1}^\infty(\lambda_l\,\sigma_l^z\sigma_{l+1}^z
+\sigma_l^x)\,. 
\label{h} 
\end{equation}  
The ${\bf\sigma}$s are Pauli spin operators and the inhomogeneous
couplings $\lambda_l$ take the HvL form: 
\begin{equation}  
\lambda_l=\lambda\left(1+\frac{a_l}{l^\kappa}\right)\,,\qquad
a_l=(-1)^{f_l}a\, ,
\label{lambdak}
\end{equation}   
where $a_l$ is the aperiodic amplitude and $a$ is related to
original parameters as indicated in~(\ref{para}). $\lambda$ plays
the role of an inverse temperature and the deviation from the
critical point, $\lambda_{\rm c}\!=\!1$, is measured by
\begin{equation}
t=1-\lambda^{-2}\,,
\label{t}
\end{equation}
so that $0\!<\!t\!<\!1$ in the ordered phase.

Using the Jordan-Wigner transformation~\cite{jordan28} and a
ca\-no\-ni\-cal transformation of the fermion operators, the Hamiltonian
can be put into diagonal form~\cite{pfeuty70,lieb61}:
\begin{equation}
{\cal H}=\sum_\alpha\epsilon_\alpha
\left(\eta_\alpha^\dag\eta_\alpha-\frac{1}{2}\right)\,,
\label{hdiag}
\end{equation}
where $\eta_\alpha^\dag$ ($\eta_\alpha$) creates (destroys) a
fermionic excitation. The excitation energies
$\epsilon_\alpha$ are obtained by solving the following
eigenvalue problem: 
\begin{equation}
\lambda_{l-1}\phi_\alpha(l\!-\!1)+(1+\lambda_{l-1}^2)\phi_\alpha(l)
+\lambda_l\phi_\alpha(l\!+\!1)=\epsilon_\alpha^2\phi_\alpha(l)\,.
\label{eigenval}
\end{equation}
The physical properties of the system can be expressed in terms of
the excitations $\epsilon_\alpha$ and the eigenvectors 
$\boldsymbol{\phi}_\alpha$ which are assumed to be normalized in the following.

\subsection{Surface magnetization}
\label{sec:2.2}

The surface magnetization $m_{\rm s}$ is given by the
matrix element $\langle0\vert\sigma_1^z\vert\sigma\rangle$,
where $\vert0\rangle$ is the vacuum state and
$\vert\sigma\rangle$ is the lowest one-particle state
$\eta_1^\dag\vert0\rangle$. Expressing $\sigma_1^z$ in terms
of diagonal fermion operators, one obtains $m_s\!=\!\phi_1(1)$, which
can be simply evaluated noticing that the first excitation
$\epsilon_1\!=\!0$ in the ordered phase~\cite{peschel84}. This leads
to:
\begin{equation}
m_{\rm s}=S^{-1/2}\,,\qquad
S=1+\sum_{j=1}^\infty\prod_{l=1}^j\lambda_l^{-2}\,. 
\label{ms}
\end{equation}
Using (\ref{lambdak}), the sum can be rewritten as:
\begin{equation}
S=1+\sum_{j=1}^\infty \lambda^{-2j}P_j^{-2}\,,\quad
P_j=\prod_{l=1}^j\left[1+a\,\frac{(-1)^{f_l}}{l^\kappa}\right]\,.
\label{pj}
\end{equation}
Since the critical behaviour is governed by the long distance
properties of the sytem, when $\kappa\!>\!0$, one may write:
\begin{eqnarray}
\ln P_j&=&\sum_{l=1}^j\ln\left[1+a\,\frac{(-1)^{f_l}}{l^\kappa}\right]
\nonumber\\
&\simeq&\sum_{l=1}^j\left[a\,\frac{(-1)^{f_l}}{l^\kappa}-
\frac{a^2}{2l^{2\kappa}}+(-1)^{f_l}O(l^{-3\kappa})\right].
\label{lnpj}
\end{eqnarray}
To go further we need some approximate expression for $f_l$. 
Eqs.~(\ref{nl1}) and~(\ref{aav}) suggest that, inside a sum, 
$f_l$ may be replaced  by
\begin{equation}
f_l\simeq \rho_\infty+c(l)\, l^{\omega-1}\,,
\label{fl}
\end{equation}
where $c(l)$ is a log-periodic amplitude. Making use of the
identity~(\ref{ident}), with $\rho_\infty\!=\!1/2$, the first sum in
the last equation of~(\ref{lnpj}) can be rewritten as: 
\begin{equation}
\sum_{l=1}^j\frac{(-1)^{f_l}}{l^\kappa}\simeq-2\sum_{l=1}^jc(l)\,
l^{\omega-\kappa-1}=-2c_\kappa(j)\sum_{l=1}^jl^{\omega-\kappa-1}\,. 
\label{suml} 
\end{equation} 
In the following, we assume that 
\begin{equation}
c_\kappa(j)=\frac{\sum_{l=1}^j c(l)\, 
l^{\omega-\kappa-1}}{\sum_{l=1}^j l^{\omega-\kappa-1}}=
\frac{\sum_{l=1}^j (f_l-1/2) 
l^{-\kappa}}{\sum_{l=1}^j l^{\omega-\kappa-1}}
\label{ckj}
\end{equation}
can be replaced by some constant effective value $\overline{c}$
when $j\gg1$ (see Appendix A for a discussion of the asymptotic
properties of $c_\kappa(j)$).  

Finally, when $\rho_\infty\!=\!1/2$ and $\kappa\!>\!0$, one obtains:
\begin{equation}
\ln P_j\simeq -\sum_{l=1}^j\left[
\frac{2\overline{c}a}{l^{1-\omega+\kappa}}
+\frac{a^2}{2l^{2\kappa}}+O(l^{-3\kappa+\omega-1})\right],
\label{lnpj2}
\end{equation}
where the size of the leading omitted term follows from a comparison 
to the first one. Otherwise, when $\overline{a}_l\neq0$, 
an additional term of order $l^{-\kappa}$ gives
the leading contribution to the sum since $\omega\!<\!1$. As mentioned 
previously, the critical behaviour would then be the same as 
for the HvL model.

\subsubsection{Relevant perturbations}

We first consider the case of relevant perturbations which,
according to the criterion of Section~\ref{sec:1}, corresponds to
$\kappa\!<\!\omega$. 

In Eq.~(\ref{lnpj2}) the first
term dominates when $1-\omega+\kappa\!<\!2\kappa$, i.e., when
$1-\omega\!<\!\kappa\!<\!\omega$, which can be satisfied only when
$\omega\!>\!1/2$. Details about the calculation of the
$t$-dependence of $m_{\rm s}$ and its finite-size  behaviour at
$\lambda_{\rm c}$ are given in Appendix B. 

The surface critical behaviour depends on the sign of
$\overline{c}a$. When $\overline{c}a\!>\!0$, the transition is 
continuous and the surface magnetization displays an essential
singularity as a function of $t$: 
\begin{eqnarray}
m_{\rm s}(t)&\sim&\exp\left[-C\, 
t^{-(\omega-\kappa)/(\kappa-\omega+1)}\right]\,,\nonumber\\
C&=&\frac{\kappa-\omega+1}{2(\omega-\kappa)}
(\overline{c}a)^{1/(\kappa-\omega+1)}\,.
\label{mst1}
\end{eqnarray}
The finite-size behaviour at the critical point is the following:
\begin{equation}
m_{\rm s,c}(L)\sim\exp
\left[-\frac{2\overline{c}a}{\omega-\kappa}\, 
L^{\omega-\kappa}\right]\,. 
\label{msl1} 
\end{equation}
When $\overline{c}a\!<\!0$ the surface remains ordered at the
bulk  critical point, the transition is first-order and the surface 
magnetization approaches its critical value linearly in $t$.

The two terms in~(\ref{lnpj2}) give contributions of the same
order to the sum when $\kappa\!=\!1-\omega$. As shown in Appendix B,
when $\kappa\!<\!1/2$ the surface critical behaviour is
modified and depends on the sign of $4\overline{c}a+a^2$.
When $\overline{c}a\!<\!-4\overline{c}^2$ or
$\overline{c}a\!>\!0$, the surface magnetization vanishes continuously
with an essential singularity:
\begin{equation}
m_{\rm s}(t)\sim\exp\left[-\frac{\kappa}{1-2\kappa}
(4\overline{c}a+a^2)^{1/2\kappa}\,
t^{1-1/2\kappa}\right]\,,
\label{mst2}
\end{equation}
and:
\begin{equation}
m_{\rm s,c}(L)\sim\exp
\left[-\frac{4\overline{c}a+a^2}{2(1-2\kappa)}\,
L^{1-2\kappa}\right]\,. 
\label{msl2}
\end{equation}
When $-4\overline{c}^2\!<\!\overline{c}a\!<\!0$, as above the
surface is ordered at the bulk critical point and the
surface transition is first-order. 

The second term of the sum in~(\ref{lnpj2}), involving $a^2$, 
is  the dominant one if $\kappa\!<\!1-\omega$. As shown in Appendix B, 
this term becomes dangerous as soon as $\kappa\!<\!1/2$. When 
$\kappa\!>\!0$, it leads to an essential singularity for the surface 
magnetization with: 
\begin{equation}
m_{\rm s}(t)\sim\exp\left[-\frac{\kappa\vert
a\vert^{1/\kappa}}{1-2\kappa}\, t^{1-1/2\kappa}\right]
\label{mst3}
\end{equation}
and
\begin{equation}
m_{\rm s,c}(L)\sim\exp\left[-\frac{a^2}{2(1-2\kappa)}\, 
L^{1-2\kappa}\right]\,.
\label{msl3}
\end{equation}
The case $\kappa\!=\!1/2\!<\!1-\omega$, where the dangerous term
leads to a marginal behaviour, will be examined in the next section.

One may notice that the surface critical behaviour is modified
by the term in $a^2$ even when $\omega\!<\!\kappa\leq1/2$, i.e.,
when the perturbation is irrelevant according to the 
criterion~(\ref{scala1}). Clearly this relevance-irrelevance 
criterion only con\-cerns the contribution of the first term 
which is linear in $a$. 

\subsubsection{Marginal perturbations} 

According to the linear criterion of 
Section~\ref{sec:1}, the aperiodic HvL perturbation is expected 
to be marginal when $\kappa\!=\!\omega$. But we shall see that this 
is the actual behaviour only when $\omega\geq1/2$. 

For $\kappa\!=\!\omega\!>\!1/2$, the first term of the sum
in~(\ref{lnpj2}) is the dominant one.
When $\overline{c}a\!>\!-1/4$, as shown in Appendix B, 
this term leads to a
second-order surface transition where $m_{\rm s}$ vanishes as
$t^{\beta_{\rm s}}$ or, at the critical point, as $L^{-x_{\rm
m}^{\rm s}}$. Since $\nu\!=\!1$, here and in what follows, both
exponents have the same expression: 
\begin{equation}
\beta_{\rm s}=x_{\rm m}^{\rm s}=\frac{1}{2}(1+4\overline{c}a)\,,
\label{betas1}
\end{equation}
When $\overline{c}a\!<\!-1/4$, the transition is first-order
and the approach to the critical surface magnetization is governed
by the exponents ${\beta'}_{\rm s}\!=\!{x'}_{\rm m}^{\rm s}$ with
\begin{equation}
\beta'_{\rm s}=-2\beta_{\rm s}(a)=-1-4\overline{c}a\,,
\label{betas2}
\end{equation}
where $\beta_{\rm s}(a)$ is the continuation of 
$\beta_{\rm s}$ in the first-order region.
\begin{figure}[t]
\epsfxsize=9cm
\begin{center}
\vglue0.mm
\hspace*{-2.mm}\mbox{\epsfbox{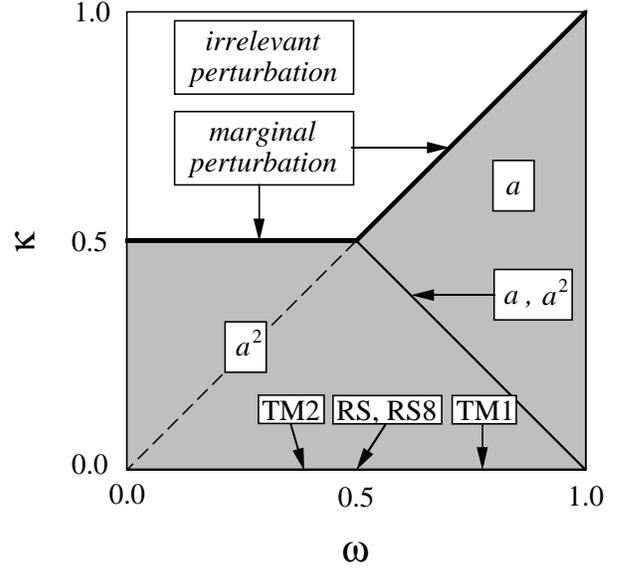}}
\end{center}
\vglue-2mm
\caption{Relevance-irrelevance of aperiodic extended surface 
perturbations in the Ising quantum chain. The perturbation is relevant
in the shaded region. To first order in the amplitude $a$, 
the perturbation is relevant only below the first diagonal. 
Below the second diagonal and the line $\kappa\!=\!1/2$ the second-order 
term is relevant and dominates the first-order one. On the marginal 
border line, the magnetization exponent is a function of $a$ above 
$\omega\!=\!1/2$, of $a^2$ below, and involves both terms at $\omega\!=\!1/2$.
The values of $\omega$ for the aperiodic sequences used in the 
numerical FSS study are indicated.}   
\label{fig-kappa-omega-1}  
\end{figure}

For $\kappa\!=\!\omega\!=\!1/2$, the two terms in~(\ref{lnpj2})
are of the same order $l^{-1}$. 
Thus one obtains a second-order surface transition with
\begin{equation}
\beta_{\rm s}=\frac{1}{2}(1+4\overline{c}a+a^2)
\label{betas3}
\end{equation}
for values of $a$ leading to a non-negative expression, i.e., when 
$\overline{c}^2\!<\!1/4$ for any $a$ and when $\overline{c}^2\!>\!1/4$ 
for $a\!<\!-2\overline{c}\!-\!\Delta$ 
or $a\!>\!-2\overline{c}\!+\!\Delta$ where 
$\Delta\!=\!(4\overline{c}^2\!-\!1)^{1/2}$. 

When $\overline{c}^2\!>\!1/4$, there is a first-order surface
transition on the interval $-2\overline{c}\!-\!
\Delta\!<\!a\!<\!-2\overline{c}\!+\!\Delta$
where $m_{\rm s}$ approaches its critical value as a power of
$t$ or $L^{-1}$ with an exponent
\begin{equation}
\beta'_{\rm s}=-2\beta_{\rm s}(a)=-1-4\overline{c}a-a^2\,. 
\label{betas4}
\end{equation}

As mentioned above, the second term in~(\ref{lnpj2}) is the
dominant one and leads to a stretched exponential surface
magnetization in the region $\kappa\!<\!1-\omega$,
$\kappa\!<\!1/2$. The marginal line
$\kappa\!=\!\omega$ of the linear relevance-irrelevance
criterion is actually moved to $\kappa\!=\!1/2$ when 
$\omega\!<\!1/2$.
On this line, the surface transition is second-order with
(see Appendix B): 
\begin{equation}
\beta_{\rm s}=\frac{1}{2}(1+a^2)\,.
\label{betas5}
\end{equation}

The behaviour in the $(\kappa,\omega)$-plane is recapitulated in 
Fig.~\ref{fig-kappa-omega-1}. Notice that on the line $\kappa\!=\!0$
the bulk is homogeneously aperiodic and, according to the Luck criterion 
in Eq.~(\ref{scala1}), the perturbation is marginal at $\omega\!=\!0$.
This marginal behaviour has been indeed confirmed by various exact 
results for aperiodic Ising models~\cite{turban94,igloi97}).
  
In Fig.~\ref{fig-kappa-omega-1} this point belongs to the domain of 
relevant perturbations due to terms of order $a^2$. This only means 
that the non-decaying aperiodicity changes the critical coupling 
from $\lambda_{\rm c}\!=\!1$ to 
$\lambda_{\rm c}\!=\!(1-a^2)^{-1/2}$\cite{pfeuty79}. At the unperturbed
fixed point value of $\lambda$, the aperiodic system is in its 
disordered phase. Thus due to terms of order $a^2$, the aperiodicity induces 
a flow towards the trivial fixed point where the coupling vanishes.

One may notice that the methods used above do not apply to the case of
hommogeneous aperiodic systems: the corresponding equations, (\ref{mst3}) 
and~(\ref{msl3}), are valid only when $\kappa\!>\!0$. 

\section{Surface magnetization of the Ising quantum chain:
finite-size-scaling study}
\label{sec:3}

The results obtained in the last section rely on the validity of the 
assumptions used to transform Eq.~(\ref{lnpj}) into Eq.~(\ref{lnpj2}). 
In this Section these results are checked through a 
numerical FSS study of the critical surface magnetization for different 
systems corresponding to the different regions of the 
$(\kappa,\omega)$-plane in Fig.~\ref{fig-kappa-omega-1}. We also 
evaluate numerically the effective amplitude $\overline{c}$ which 
is needed when the term linear in $a$ contributes to the critical 
behaviour.
\begin{figure}[t]
\epsfxsize=9cm
\begin{center}
\vglue0.mm
\hspace*{-2.mm}\mbox{\epsfbox{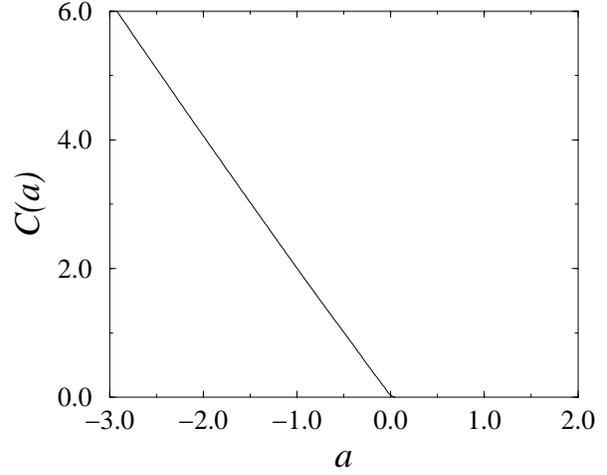}}
\end{center}
\vglue-7mm
\caption{Amplitude $C(a)$ for the TM1 sequence 
with $\kappa\!=\!1/2$ deduced from FSS at criticality with $L_n\!=\!6^n$, 
$n\!=\!1,7$. The surface transition is 
continuous for $a\!<\!0$ and first-order for $a\!>\!0$.}   
\label{fig-tm1-k05}  
\end{figure}
\begin{figure}[ht]
\epsfxsize=9cm
\begin{center}
\vglue0.mm
\hspace*{-2.mm}\mbox{\epsfbox{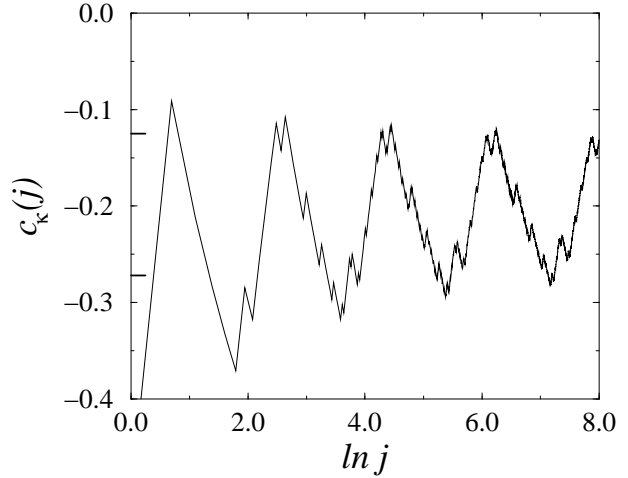}}
\end{center}
\vglue-7mm
\caption{Amplitude $c_\kappa(j)$ as a function of $\ln j$ 
for the TM1 sequence with $\kappa\!=\!1/2$. Asymptotically $c_\kappa(j)$
oscillates log-periodically between the two values indicated on 
the left axis.}  
\label{fig-tm1-c-k05}  
\end{figure}
\begin{figure}[ht]
\epsfxsize=9cm
\begin{center}
\vglue0.mm
\hspace*{-2.mm}\mbox{\epsfbox{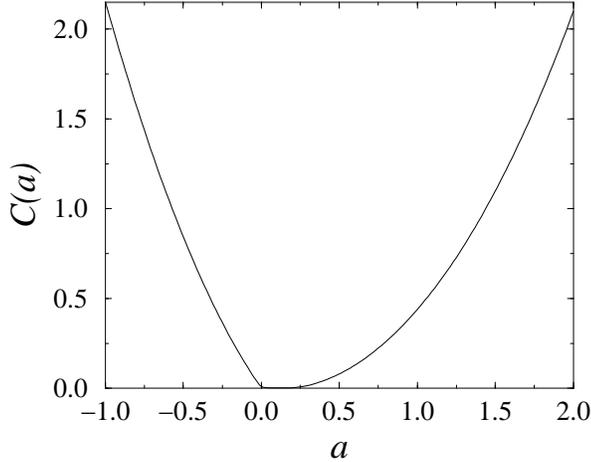}}
\end{center}
\vglue-7mm
\caption{Amplitude $C(a)$ for the TM1 sequence with 
$\kappa\!=\!1-\omega$ deduced from FSS at criticality with $L_n\!=\!6^n$, 
$n\!=\!1,5$. The surface transition is 
first-order in the intermediate region where $C(a)$ vanishes 
and continuous elsewhere.}   
\label{fig-tm1-k1-omega}  
\end{figure}
\subsection{Aperiodic sequences}
We used aperiodic sequences with the same asymptotic density 
$\rho_{\infty}\!=\!1/2$ leading to different values of the wandering 
exponent $\omega$ and also to different behaviours for the log-periodic
amplitude $c_{\kappa}(j)$.

The Rudin-Shapiro (RS) sequence~\cite{dekking83} follows from
substitutions on the letters $A$, $B$, $C$, $D$, with 
$S(A)\!=\!AB$, $S(B)\!=\!AC$, $S(C)\!=\!DB$, $S(D)\!=\!DC$.
The substitution matrix has eigenvalues $\Lambda_1\!=\!2$ and 
$\Lambda_2\!=\!\sqrt{2}$ so that $\omega\!=\!1/2$. The different letters have 
the same asymp\-to\-tic density $\rho_\infty(i)\!=\!1/4$ $(i\!=\!A$, $B$, 
$C$, $D)$. Each letter in the sequence corresponds to a pair of digits 
$A\!=\!00$, $B\!=\!01$, $C\!=\!10$ and $D\!=\!11$, which gives $\rho_\infty\!=\!1/2$.

The same substitutions on the letters are used to generate the 
RS8 sequence which is obtained by replacing 0 and 1 by 0000 and 1111 in the 
the RS sequence. The substitution matrix being the same, both the 
wandering exponent and the asymptotic density keep their
RS values. The new correspondance between letters and digits only 
affect the behaviour of the log-periodic amplitude $c_{\kappa}(j)$.

The RS and RS8 sequences are self-similar under dilatation 
by a factor $b\!=\!4$.  

Other values of the wandering exponent have been obtained through 
decoration of the Thue-Morse sequence which follows from the 
substitutions $S(0)\!=\!01$ and $S(1)\!=\!10$~\cite{dekking83}. 

The  TM1 sequence is generated through $S(0)\!=\!010000$ and 
$S(1)\!=\!101111$. The two eigenvalues of the substitution matrix are 
$\Lambda_1\!=\!6$ and $\Lambda_2\!=\!4$ so that $\omega\!=\!\ln 4/\ln 6\!\simeq\! 
0.7737$. Due to the invariance under the exchange of 0 and 1, the two 
digits have the same asymp\-to\-tic density $\rho_\infty\!=\!1/2$.

Finally $S(0)\!=\!010100$ and $S(1)\!=\!101011$ lead to 
the TM2 sequence. Its substitution matrix has eigenvalues 
$\Lambda_1\!=\!6$ and $\Lambda_2\!=\!2$, which gives $\omega\!=\!\ln 2/\ln 
6\!\simeq\! 0.3869$. For the same reason as above $\rho_\infty\!=\!1/2$.

The TM1 and TM2 sequences are self-similar under dilatation by $b\!=\!6$.

\subsection{Relevant perturbations}
\begin{figure}[t]
\epsfxsize=9cm
\begin{center}
\vglue0.mm
\hspace*{-2.mm}\mbox{\epsfbox{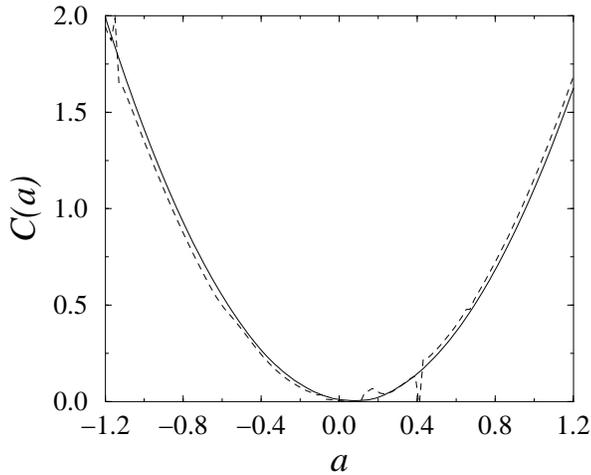}}
\end{center}
\vglue-7mm
\caption{Amplitude $C(a)$ for the RS sequence with 
$\kappa\!=\!0.3$ deduced from FSS at criticality with $L_n\!=\!4^n$, 
$n\!=\!1,9$. The surface magnetization vanishes continuously
with a parabolic amplitude in the stretched exponential. 
The solid (dashed) line corresponds to a correction-to-scaling 
exponent $0.5$ ($0.01$) in the BST extrapolation.}   
\label{fig-rs-k03-2fits}  
\end{figure}
\begin{figure}[ht]
\epsfxsize=9cm
\begin{center}
\vglue0.mm
\hspace*{-2.mm}\mbox{\epsfbox{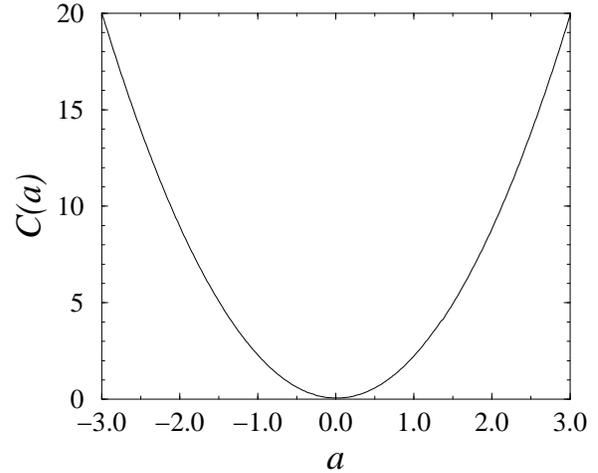}}
\end{center}
\vglue-7mm
\caption{Amplitude $C(a)$ for the TM2 sequence with 
$\kappa\!=\!\omega$ deduced from FSS at criticality with 
$L_n\!=\!6^n$, $n\!=\!1,7$. The surface magnetization vanishes continuously with a 
pa\-ra\-bo\-lic amplitude in the stretched exponential.}   
\label{fig-tm2-komega}  
\end{figure}
In the case of relevant perturbations, according to~(\ref{msl4}), 
$\ln m_{\rm s,c}$ is linear in $L^\tau$ with a slope $-C(a)$ 
when the transition is continuous. We studied the size-dependence
of $\ln m_{\rm s,c}$ for sizes of the form $L_n\!\sim\!b^n$. 
Approximants $C_n(a)$ for the amplitude, deduced from two-point 
fits for successive sizes $L_n$, $L_{n+1}$, were extrapolated using the BST 
algorithm~\cite{henkel88}. When the transition is first-order, the 
leading contribution is independent of the size of the system and 
$C(a)$ vanishes. 

With the TM1 sequence at $\kappa\!=\!1/2$, the critical behaviour 
is governed by a term of order $a$ (see Fig.~\ref{fig-kappa-omega-1}). 
The extrapolated amplitude $C(a)$ is shown in Fig.~\ref{fig-tm1-k05}. 
It displays the expected linear variation with $a$ in the region $a\!<\!0$ 
where the transition is continuous. The slope leads to the effective 
amplitude $\overline{c}$ given in Table~1.

Since $\kappa\!<\!\omega$ the amplitude $c_\kappa(j)$, which is shown 
in Fig.~\ref{fig-tm1-c-k05}, remains log-periodic at infinity 
(see Appendix A). For $a\!<\!0$ the main contribution to $P_j^{-2}$ 
in~(\ref{b1}) comes, at large $j$-values, from the absolute minima 
of $c_\kappa(j)$ and $\overline{c}$ in Table~1 corresponds 
to the extrapolated value $(c_{\kappa})_{\rm min}$.

\begin{table*}[t]\sidecaption 
\protect\label{tab:1}       
\begin{tabular}{cccccc}
\hline\noalign{\smallskip}
sequence & TM1 & TM1 $(a\!<\!0)$ & TM1 $(a\!>\!a_{\rm 
c})$ & RS & TM2 \\
\noalign{\smallskip}\hline\noalign{\smallskip}
$\omega$ & $0.7737$  & $0.7737$ & $0.7737$ & $1/2$ & $0.3869$ \\
$\kappa$ & $1/2$ & $1-\omega$ & $1-\omega$ & $0.3$ & $\omega$ \\
$\overline{c}$ from $c_\kappa(j)$ & $-0.272(7)$ & $-0.34(1)$ & 
$-0.055(1)$ &---&---\\
$\overline{c}$ from $C(a)$ & $-0.271(5)$ & $-0.332(3)$ &-0.04(3)&---&---\\
$\gamma_2$ (expected) &---& $0.9134$ & $0.9134$ & $5/4$ & $2.20951$\\
$\gamma_2$ (numerical) &---& $0.92(2)$ & $0.6(1)$ & $1.251(1)$ & 
$2.209(1)$ \\
\noalign{\smallskip}\hline
\end{tabular}
\caption{Parameters of the amplitude $C(a)$ in the stretched 
exponential deduced from the FSS study 
of the surface mag\-ne\-ti\-za\-tion at cri\-ti\-ca\-li\-ty. $\gamma_2$ is the 
coefficient of the quadratic term in $C(a)$.}
\end{table*}

For the TM1 sequence with $\kappa\!=\!1-\omega$ the surface 
magnetization at criticality is given by~(\ref{msl2}). In 
Fig.~\ref{fig-kappa-omega-1}, this system is on the solid line 
inside the domain of relevant perturbations where both $a$ and $a^2$ 
contribute to the critical behaviour.

The FSS study leads to the extrapolated amplitude $C(a)$ 
shown in Fig.~\ref{fig-tm1-k1-omega}. We were limited to sizes up to 
$6^5$ because the exponent of $L$ in the stretched exponential is two times 
larger than before and $m_{\rm s,c}(L)$ becomes quite small at large size. 
The corresponding parameters are given in Table~1. 

As above $c_\kappa(j)$ remains asymptotically oscillating, here between
$(c_{\kappa})_{\rm min}\!=\!-0.34(1)$ and $(c_{\kappa})_{\rm 
max}\!=\!-0.055(1)$.

When $a\!<\!0$ the transition is continuous and $C(a)$ displays the 
expected parabolic variation. The coefficient of the linear term 
leads to an effective amplitude $\overline{c}$ in good agreement 
with the value of $(c_{\kappa})_{\rm min}$ which gives the main 
contribution to $P_j^{-2}$ in~(\ref{b2}). 

When $a\!>\!0$ the transition is first-order below a critical value
$a_{\rm c}$. In the region $a\!>\!a_{\rm c}$ where the 
transition is con\-ti\-nu\-ous, $\overline{c}$ is close to 
$(c_{\kappa})_{\rm max}$ as expected but the agreement is poor for the
coefficient of the quadratic term. In both cases the sizes used in the 
FSS study are too small to obtain truly reliable estimates. 

The RS sequence with $\kappa\!=\!0.3$ leads to a relevant perturbation 
for which the term in $a^2$ governs the critical behaviour as 
indicated in Eq.~(\ref{msl3}). The extrapolated amplitude $C(a)$ is shown in 
Fig.~\ref{fig-rs-k03-2fits} and the parameters given in Table~1. 
\begin{figure}[t]
\epsfxsize=9cm
\begin{center}
\vglue0.mm
\hspace*{-2.mm}\mbox{\epsfbox{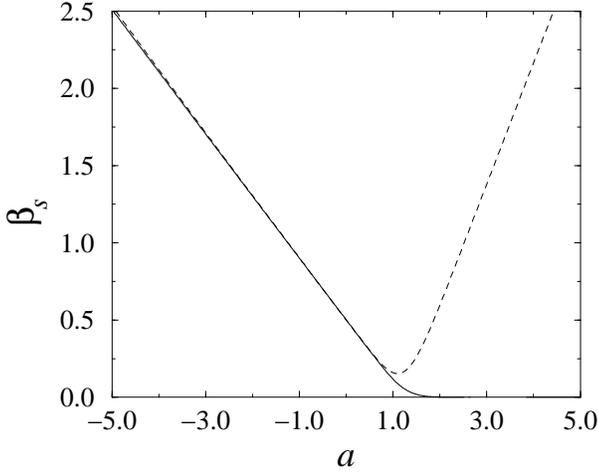}}
\end{center}
\vglue-7mm
\caption{Surface magnetization exponents for the TM1 sequence 
with $\kappa\!=\!\omega$. The solid line corresponds to 
$\beta_{\rm s}$ at the second-order surface transition and the dashed 
line, in the first-order region on the right, to $\beta'_{\rm s}$ 
which controls the approach to the critical 
magnetization.}   
\label{fig-tm1-komega}  
\end{figure}
\begin{figure}[ht]
\epsfxsize=9cm
\begin{center}
\vglue0.mm
\hspace*{-2.mm}\mbox{\epsfbox{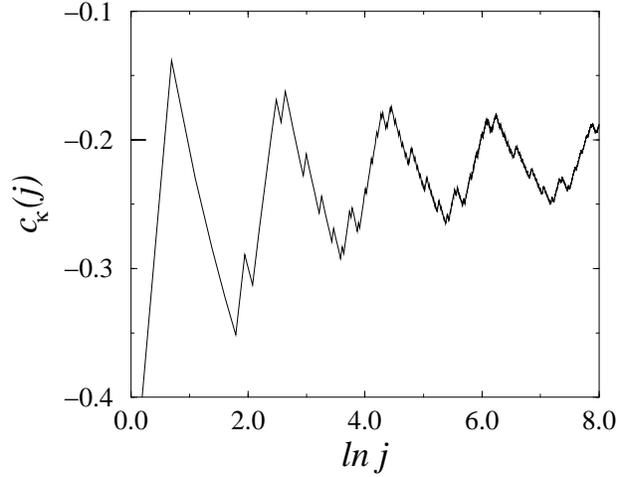}}
\end{center}
\vglue-7mm
\caption{Amplitude $c_\kappa(j)$ as a function of $\ln j$ 
for the TM1 sequence with $\kappa\!=\!\omega$. Asymptotically $c_\kappa(j)$
converges towards the effective amplitude $\overline{c}$ indicated on 
the left axis.}   
\label{fig-tm1-c-komega}  
\end{figure}

For the solid line, obtained with a 
correction-to-scaling exponent equal to $0.5$ in the BST extrapolation, 
the parabola is not centered on $a\!=\!0$: there is a weak linear 
contribution to the amplitude. With a 
correction-to-scaling exponent equal to $0.01$ (dashed line in 
Fig.~\ref{fig-rs-k03-2fits}) the extrapolation is less stable 
but the coefficient of the linear term is reduced from 
$-0.15$ to $-0.09$. Thus we suspect that the unexpected linear 
contribution to $C(a)$ is a correction-to-scaling effect. 
One may notice that a power of $L$ in front of the stretched 
exponential leads to a logarithmic correction.

Next we consider the TM2 sequence with $\kappa\!=\!\omega$. The 
corresponding point in the $(\kappa,\omega)$-plane 
belongs to the dashed line in 
Fig.~\ref{fig-kappa-omega-1}. It leads to a perturbation which 
is marginal to linear order in $a$ but becomes relevant due to the 
term of order $a^2$. The extrapolated amplitude 
is shown in Fig.~\ref{fig-tm2-komega}.

The amplitude $C(a)$ is still given by Eq.~(\ref{msl3}) and displays a 
parabolic behaviour. The coefficient $\gamma_2$ in Table~1 is in 
goog agreement with the expected value. The same parabolic 
variation was obtained for the TM2 sequence with 
$\kappa\!=\!0.3$ and $\kappa\!=\!0.4$.

\subsection{Marginal perturbations} 
\begin{figure}[t]
\epsfxsize=9cm
\begin{center}
\vglue0.mm
\hspace*{-2.mm}\mbox{\epsfbox{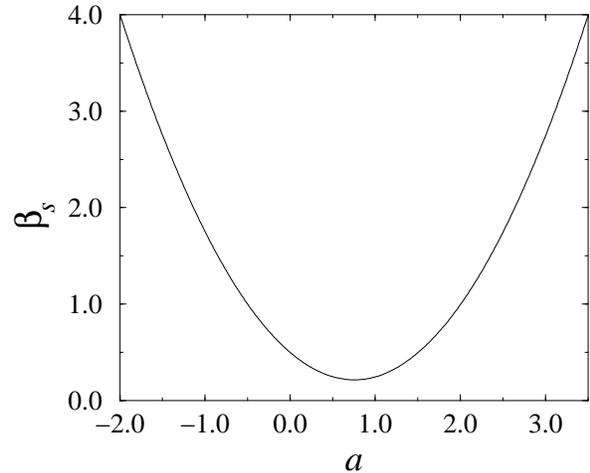}}
\end{center}
\vglue-7mm
\caption{Surface magnetization exponent $\beta_{\rm s}$ for 
the RS sequence with $\kappa\!=\!\omega$. The surface transition is 
always second-order.}   
\label{fig-rs-komega}  
\end{figure}
\begin{figure}[ht]
\epsfxsize=9cm
\begin{center}
\vglue0.mm
\hspace*{-2.mm}\mbox{\epsfbox{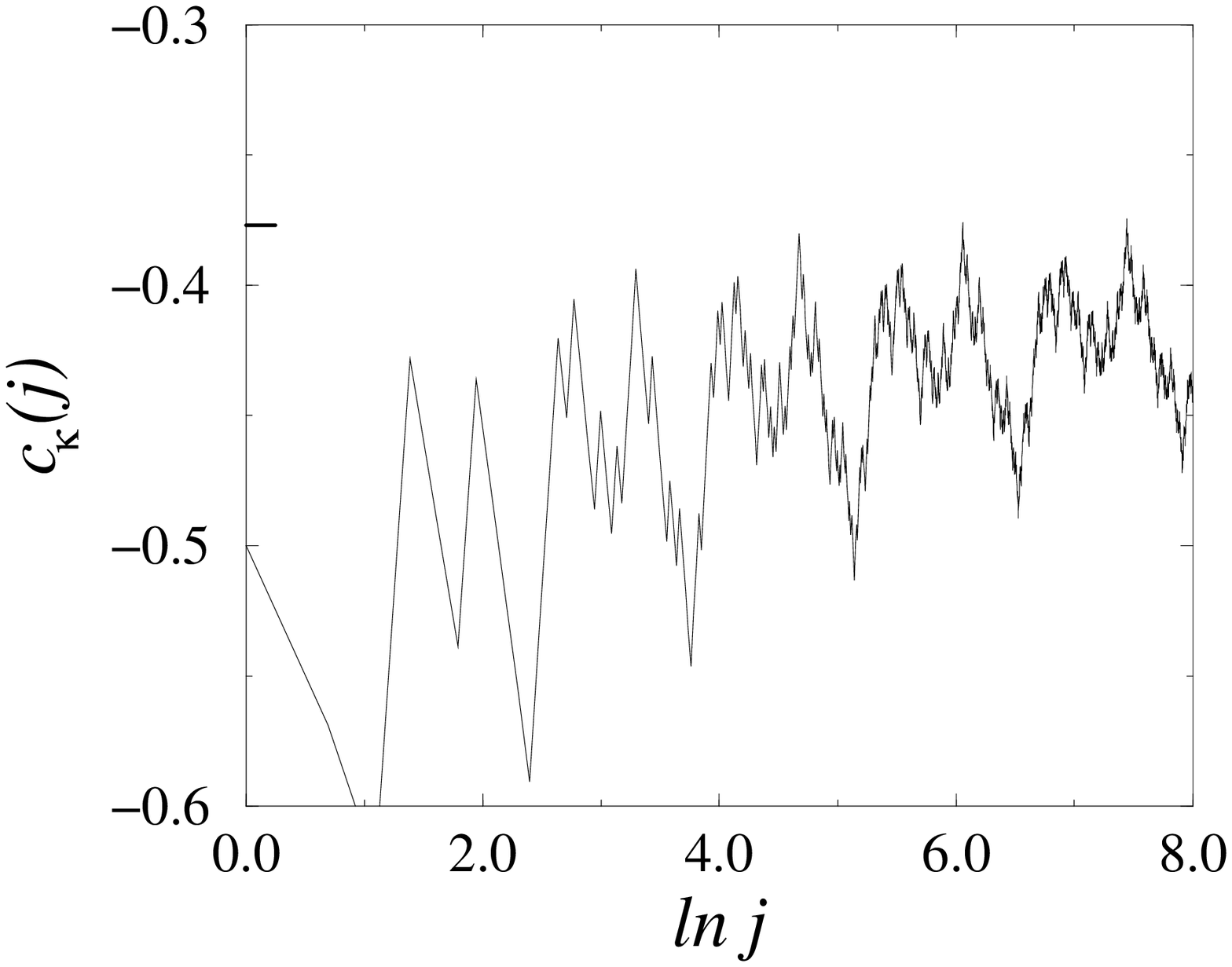}}
\end{center}
\vglue-7mm
\caption{Amplitude $c_\kappa(j)$ as a function of $\ln j$ 
for the RS sequence with $\kappa\!=\!\omega$. Asymptotically $c_\kappa(j)$
converges towards the effective amplitude $\overline{c}$ indicated on 
the left axis.}   
\label{fig-rs-c-komega}  
\end{figure}
\begin{figure}[t]
\epsfxsize=9cm
\begin{center}
\vglue0.mm
\hspace*{-2.mm}\mbox{\epsfbox{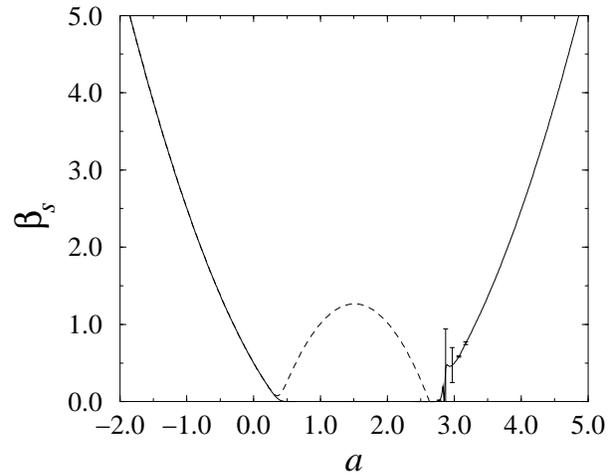}}
\end{center}
\vglue-7mm
\caption{Surface magnetization exponents for 
the RS8 sequence with $\kappa\!=\!\omega$. The solid line corresponds to 
$\beta_{\rm s}$ at the second-order surface 
transition and the dashed line to $\beta'_{\rm s}$ 
which controls the approach to the critical 
magnetization. The chain sizes used in the FSS study are of 
the form $L_n\!=\!2\times4^n$, $n\!=\!1,8$, with one size less for 
$\beta'_{\rm s}$.}   
\label{fig-rs8-komega}  
\end{figure}
\begin{figure}[ht]
\epsfxsize=9cm
\begin{center}
\vglue0.mm
\hspace*{-2.mm}\mbox{\epsfbox{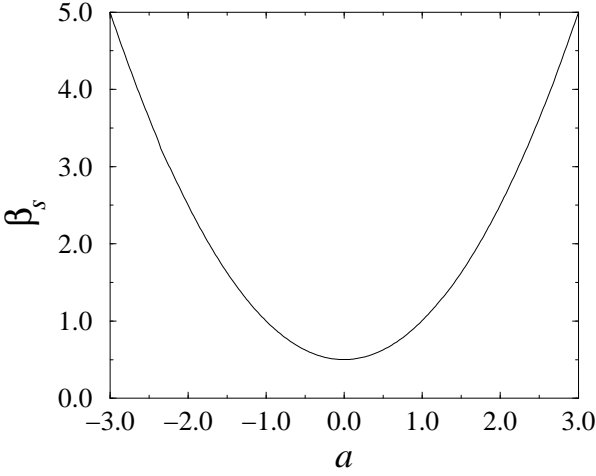}}
\end{center}
\vglue-7mm
\caption{Surface magnetization exponent $\beta_{\rm s}$ for 
the TM2 sequence with $\kappa\!=\!1/2$. The transtion is always second-order
and the critical behaviour is governed by $a^2$.}   
\label{fig-tm2-k05}  
\end{figure}
\begin{table*}[t]\sidecaption
\label{tab:2}       
\begin{tabular}{ccccc}
\hline\noalign{\smallskip}
sequence & TM1 & RS & RS8 & TM2 \\
\noalign{\smallskip}\hline\noalign{\smallskip}
$\omega$ & $0.7737$  & $1/2$ & $1/2$ & $0.3869$ \\
$\kappa$ & $\omega$ & $\omega$ & $\omega$ & $1/2$ \\
$\overline{c}$ from $c_\kappa(j)$ & $-0.199(1)$ & $-0.377(4)$ & 
$-0.75(1)$ &---\\
$\overline{c}$ from $\beta_{\rm s}(a)$ & $-0.1998(1)$ & $-0.3763(1)$ & 
$-0.7526(2)$ &---\\
$\overline{c}$ from $\beta'_{\rm s}(a)$ & $-0.19(1)$ &---& 
$-0.7525(15)$ &---\\
$\beta_0$ & $0.49998(2)$ & $0.4997(3)$ & $0.4999(5)$ & $0.502(5)$\\
$\beta_2$ &---& $0.50007(7)$ & $0.5000(1)$ & $0.499(1)$\\
$\beta'_0$ & $-0.99(2)$ &---& $-0.999(1)$ &---\\
$\beta'_2$ &---&---& $-0.999(1)$ &---\\
\noalign{\smallskip}\hline
\end{tabular}
\caption{Parameters of the magnetic exponents $\beta_{\rm s}(a)$
and $\beta'_{\rm s}(a)$ deduced from the FSS study 
of the surface magnetization at criticality. The expected value for 
the constant term $\beta_0$ ($\beta'_0$) and the coefficient of the 
quadratic term $\beta_2$ ($\beta'_2$) in $\beta_{\rm s}(a)$ is $1/2$
($-1$).}
\end{table*}

The critical surface magnetization has 
been calculated using the same chain sizes as in the relevant 
situation (up to $n\!=\!7$ for TM1). 
Estimates for the exponent $\beta_{\rm s}$ are obtained via two-point 
fits of $\ln[m_{\rm s,c}(L_n)]$ versus $\ln L_{n}$. The two-point approximants 
are extrapolated using the BST algorithm. In the first-order regions, 
the regular contribution associated with the 
critical magnetization $m_{\rm s,c}(\infty)$ is eliminated using 
the differences $\Delta m_{\rm s,c}(L_n)\!=\!m_{\rm s,c}(L_n)-m_{\rm s,c}
(L_{n+1})$ and the usual procedure is applied to calculate 
the exponent $\beta'_{\rm s}$, although with one size less. 

As a first example of marginal behaviour we consider the TM1 
sequence with $\kappa\!=\!\omega$. The extrapolated exponent values 
are shown in Fig.~\ref{fig-tm1-komega}.

In agreement with Eqs.~(\ref{betas1}) and~(\ref{betas2}), both 
exponents vary linearly with $a$. Due to the finite sizes 
used, the singularities remain rounded near $a_{\rm c}$ 
where the surface transition changes from second to first order.

As shown in Fig.~\ref{fig-tm1-c-komega}, $c_\kappa (j)$ here converges 
towards the effective amplitude $\overline{c}$. It is compared to the 
values deduced from the slopes of the surface exponents in 
Table~2. The precision on the parameters deduced from 
$\beta'_{\rm s}(a)$ is lower since 
the number of points is reduced by one in the extrapolation.

The RS sequence with $\kappa\!=\!\omega\!=\!1/2$ leads to a marginal behaviour
with linear and quadratic contributions to the exponents. 
The exponent $\beta_{\rm s}$ is shown in Fig.~\ref{fig-rs-komega}.
The corresponding parameters are given in Table~2.
The variation is parabolic in agreement with Eq.~(\ref{betas3}) and 
the absence of a first-order region is linked to the small value of
$\vert\overline{c}\vert$. The log-periodic amplitude $c_\kappa(j)$ shown in 
Fig.~\ref{fig-rs-c-komega} converges to $\overline{c}\!=\!-0.377(4)$
whereas the limiting value for the occurence of a first-order 
transition is $\vert\overline{c}\vert\!=\!1/2$. 

With the RS8 sequence at $\kappa\!=\!\omega\!=\!1/2$ we obtain once more a 
marginal perturbation. But now $c_\kappa(j)$ converges to a value 
allowing for the occurence of a first-order transition. 
The exponents $\beta_{\rm s}$ and $\beta'_{\rm s}$ are shown in 
Fig.~\ref{fig-rs8-komega}. 

The surface transition is first-order in the 
central region and second-order outside, with a parabolic variation of 
the exponents in both cases, in agreement with the analytical 
expressions in Eqs.~(\ref{betas3}) and~(\ref{betas4}). The accuracy of 
the extrapolation is reduced near the singularities on the borders.
The coefficients given in Table~2 are in good agreement 
with the expected ones.

Finally we have studied the TM2 sequence with $\kappa\!=\!1/2$ as an 
example of a system for which the marginal behaviour is induced by the 
term of order $a^2$. In this case the surface transition is always 
second order. The exponent $\beta_{\rm s}$ in 
Fig.~\ref{fig-tm2-k05} displays the parabolic behaviour obtained  
in Eq.~(\ref{betas5}). Here too the coefficients, given in 
Table~2, are close to the expected values.

\section{Discussion}
\label{sec:4}
We have seen that the first-order relevance-irrelevance criterion of 
Section~\ref{sec:1} does not predict completely the actual influence of 
aperiodic extended surface perturbations. In some domains of the 
$(\kappa,\omega)$-plane, second-order contributions govern the 
surface critical behaviour. In order to clear this point, one has to 
look at the scaling behaviour of second-order terms in the 
perturbation expansion of the free energy. 

We shall consider the case of 
a $d$-dimensional system with a free surface at $l\!=\!0$, perpendicular to the unit 
vector $\vec{u}$. The perturbation term is given by:
\begin{equation}
-\beta V=\sum_{\vec{r}}\frac{A_{l}}{l^\kappa}\,\varepsilon(\vec{r})\,,
\label{v}
\end{equation}
where $\varepsilon(\vec{r})$ is an energy density operator which, for the Ising model, 
takes the form $\sigma_{\vec{r}}\sigma_{\vec{r}+\vec{u}}$. $A_{l}$ 
is the perturbation amplitude defined in~(\ref{k2l}). The 
position vector $\vec{r}$ will be decomposed into its components 
perpendicular and parallel to the surface as 
$l\vec{u}+\vec{r}_{\parallel}$. The second-order correction to the free 
energy is given by:
\begin{eqnarray}
-\beta F^{(2)}&=&\frac{1}{2}\beta^{2}[\langle V^{2}\rangle-\langle 
V\rangle^2]\nonumber\\
&=&\frac{1}{2}\sum_{\vec{r},\vec{r'}}\frac{A_{l}A_{l'}}{(ll')^\kappa}\,{\cal 
G}_{\varepsilon\varepsilon}(\vec{r},\vec{r'})\,,
\label{f2}
\end{eqnarray}
where $\langle\cdots\rangle$ denotes a thermal 
average and ${\cal G}_{\varepsilon\varepsilon}$ is the connected 
energy-energy 
correlation function, both for the unperturbed system. 
\begin{figure}[t]
\epsfxsize=9cm
\begin{center}
\vglue0.mm
\hspace*{-2.mm}\mbox{\epsfbox{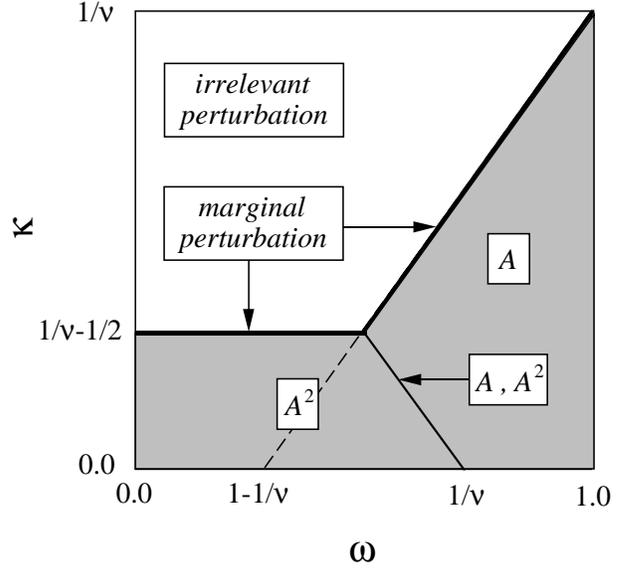}}
\end{center}
\vglue-2mm
\caption{Relevance-irrelevance of aperiodic extended surface 
perturbations in a system with bulk correlation length exponent $\nu$. 
The perturbation is relevant in the shaded region. The order of the 
dominant perturbation terms is indicated, assuming that 
off-diagonal terms do not contribute.}   
\label{fig-kappa-omega-2}  
\end{figure}

We first consider off-diagonal contributions to $F^{(2)}$ 
which come from pairs of sites belonging to different layers.
The corresponding amplitude $(A^2)_{\rm od}$ scales as 
the product of two first-order amplitudes with:
\begin{equation}
({A'}^2)_{\rm od}=b^{2(-\kappa+\omega-1+1/\nu)}(A^2)_{\rm od}\,.
\label{scala2od}
\end{equation}
One may notice that such cross-terms do not enter into the calculation 
of the surface magnetization of the $2d$ Ising model.

The diagonal contributions to $F^{(2)}$ involve pairs 
of sites belonging to the same layer and read:
\begin{equation}
-\beta F^{(2)}_{\rm d}
=\frac{L^{d-1}}{2}\sum_{l}\sum_{\vec{r}_\parallel}
\frac{A^2}{l^{2\kappa}}\,{\cal 
G}_{\varepsilon\varepsilon}(r_\parallel)\,,
\label{f2d}
\end{equation}
where $L^{d-1}$ is the surface of the layers.
The density under the sums has dimension $2d-1$ whereas the 
correlation function has dimension $2x_{\rm e}$, where $x_{\rm e}$ is 
the scaling dimension of the bulk energy density, thus one obtains:
\begin{equation}
\frac{({A'}^2)_{\rm d}}{{l'}^{2\kappa}}
=b^{2d-1-2x_{\rm e}}\frac{(A^2)_{\rm d}}{l^{2\kappa}}
=b^{-1+2/\nu}\frac{(A^2)_{\rm d}}{l^{2\kappa}}
\label{a2d}
\end{equation}
and the diagonal second-order amplitude transforms as:
\begin{equation}
({A'}^2)_{\rm d}=b^{-1-2\kappa+2/\nu}(A^2)_{\rm d}\,.
\label{scala2d}
\end{equation}
It is a relevant (irrelevant) variable when $\kappa\!<\!(>)-1/2+1/\nu$ 
and a marginal one when $\kappa\!=\!-1/2+1/\nu$.

The scaling dimension $y_{(A^2)_{\rm d}}$ of $(A^2)_{\rm d}$ has to 
be compared to the dimension $y_A$ of $A$ given in $(\ref{scala1})$ 
to see  which term governs the critical behaviour when both are relevant. 
When $y_{(A^2)_{\rm d}}\!>\!y_A$, i.e., when 
$\kappa\!<\!-\omega+1/\nu$, the second-order term dominates and one expects 
an $A^2$-dependence of the amplitudes in the stretched exponentials.
On the line $\kappa\!=\!-\omega+1/\nu$ when $\kappa\!\leq\!-1/2+1/\nu$, 
linear and quadratic terms contribute together.

A summary of the relevance-irrelevance in the $(\kappa,\omega)$-plane 
is given in Fig.~\ref{fig-kappa-omega-2}. It is in agreement 
with our findings of Sections~\ref{sec:2} and~\ref{sec:3} for the surface 
magnetization of the $2d$ Ising model. The order of the dominant 
contributions could be mo\-di\-fied by the discarded off-diagonal 
second- or higher-order terms. As mentioned earlier for the 
$2d$ Ising model, there is a shift of the bulk critical coupling for the 
homogeneous aperiodic system on the line $\kappa\!=\!0$. Along this line 
the perturbation is irrelevant below $\omega\!=\!1-1/\nu$. 

One may notice that the model shows some kind of universal 
behaviour. Both the amplitude $C(a)$ in the stretched 
exponential and the surface magnetic exponent $\beta_{\rm s}(a)$
are independent of the aperiodic sequence, provided the value of 
$\omega$ is such that the system belongs to the region $\kappa\leq 1/2$, 
$\kappa\!<\!1-\omega$ of Fig.~\ref{fig-kappa-omega-1} where the 
quadratic contribution dominates. 

Let us now briefly discuss the behaviour of the first gap and the 
surface energy. 

On a finite critical Ising chain, the first gap is known to scale 
as~\cite{igloi97}
\begin{equation}
\epsilon_1(L)\sim m_{\rm s,c}(L)\, P_L^{-1}\,\overline{m}_{\rm s,c}(L)\,, 
\label{epsi}
\end{equation}
where $\overline{m}_{\rm s,c}$ is the critical magnetization 
on the second surface and $P_{L}$ is the product of the
couplings defined in Eq.~(\ref{pj}). When $\kappa\!>\!0$ the bulk is 
unperturbed and, asymptotically, the second surface displays 
an ordinary surface transition. Thus the scaling 
dimension of $\overline{m}_{\rm s}$ is $\overline{x}{}_{\rm m}^{\rm 
s}\!=\!1/2$. 

For relevant perturbations, with the notations of Appendix B, we have:
\begin{equation}
P_L^{-1}\simeq\exp(\frac{1}{2}B\, L^\tau)\; .
\label{pl1}
\end{equation}

When the surface transition is continuous $m_{\rm s,c}(L)$,
given in Eq.~(\ref{msl4}), behaves as $P_L$. It follows that the first gap 
vanishes as a power of $L$. One expects in this case the 
unperturbed behaviour $\epsilon_1(L)\!\sim\! L^{-1}$ . 

When the surface transition is first-order, the leading term in 
$m_{\rm s,c}(L)$ is a constant and there is no more com\-pen\-sa\-tion. 
The first gap is anomalous, it vanishes with an essential singularity. 
This behaviour is linked to the localization of the corresponding 
eigenvector $\boldsymbol{\phi}_1$, which itself is 
reponsible for the finite weight on the first component $\phi_1(1)$, 
leading to a non-vanishing surface magnetization~\cite{peschel84}. 

The scaling dimension $x_{\rm e}^{\rm s}$ of the surface energy can 
be deduced from the finite-size behaviour 
$e_{\rm s}\!=\!\langle0\vert\sigma_1^{x}\vert\varepsilon\rangle$ where 
the state $\vert\varepsilon\rangle\!=\!\eta_1^\dagger\eta_2^\dagger\vert0\rangle$ 
is the lowest two-particle excitated state. This matrix element can be 
written as~\cite{karevski95}:
\begin{equation}
e_{\rm s}=(\epsilon_2-\epsilon_1)\phi_1(1)\phi_2(1)\,.
\label{es}
\end{equation}
For relevant perturbations one expects an essential singularity 
since at least $\phi_2(1)$ should behave in this way.

In the case of a marginal perturbation, with the notations of Appendix 
B, one may write:
\begin{equation}
P_L^{-1}\simeq L^{B/2}=L^{-1/2+\beta_{\rm s}(a)}\,,
\label{pl2}
\end{equation}
where $\beta_{\rm s}(a)$ is the surface magnetic exponent $\beta_{\rm 
s}$ when the transition is second-order or its continuation to negative 
values when the transition is first-order, i.e., when $\beta_{\rm 
s}\!=\!0$.
 
When the transition is second-order Eqs.~(\ref{epsi}) and~(\ref{pl2}) 
lead to:
\begin{equation}
\epsilon_1(L)\sim L^{-\beta_{\rm s}}\, L^{-1/2+\beta_{\rm s}}\, 
L^{-1/2}\sim L^{-1}\,,
\label{epsi1}
\end{equation}
i.e., to the unperturbed behaviour as above for relevant 
perturbations when the transition is continuous.

For the surface energy, Eq.~(\ref{es}) leads to the scaling dimension
\begin{equation}
x_{\rm e}^{\rm s}=1+2\beta_{\rm s}
\label{xes1}
\end{equation}
if one assumes that $\phi_2(1)$ scale as 
$L^{-\beta_{\rm s}}$ too. This is known to be true  either for the 
marginal HvL model~\cite{berche90} or for the aperiodic version of the 
same model~\cite{karevski95} where the couplings are modulated according 
to the Fredholm sequence~\cite{dekking83}. 

When the transition is first-order, due to the localization of 
$\boldsymbol{\phi}_1$, the scaling of the first excitation is 
anomalous~\cite{gap}:
\begin{equation}
\epsilon_1(L)\sim L^{-1+\beta_{\rm s}(a)}\,.
\label{epsi2}
\end{equation}
It decays faster than higher excitations since $\beta_{\rm s}(a)\!<\!0$. 

For the surface energy, we conjecture the following beha\-viour:
\begin{equation}
x_{\rm e}^{\rm s}=1+\frac{\beta'_{\rm s}}{2}=1-\beta_{\rm s}(a)\,.
\label{xes2}
\end{equation}
The factor containing the excitations in Eq.~(\ref{es}) is dominated 
by $\epsilon_2$ which vanishes as $L^{-1}$. Furthermore we assumed that, 
like in Refs.~\cite{karevski95,berche90}, $\phi_2(1)$ scales as 
$L^{-\beta'_{\rm s}/2}$.

As for the HvL model, in the regime of first-order transition, 
the anomalous scaling of the first gap leads 
to an exponent asymmetry~\cite{blote83}. 
For example, in the disordered phase, the exponent of the correlation 
length $\xi_{\parallel}$, along the surface of the semi-infinite 
system, is governed by the first gap and 
$\nu_{\parallel}\!=\!1-\beta_{\rm s}(a)$. In the ordered phase the 
first excitation vanishes and $\xi_{\parallel}\!\sim\epsilon_2^{-1}$
so that $\nu'_{\parallel}\!=\!1$, like in the unperturbed system.    

To conclude, let us mention that the case of random surface extended 
perturbations, which has fluctuation properties similar to the 
aperiodic case with $\omega\!=\!1/2$~\cite{igloi98}, is currently 
under study.

\begin{acknowledgement}
I thank Ferenc Igl\'oi and Dragi
Karevski for constructive comments and a long collaboration on 
aperiodic systems.
 
\end{acknowledgement}

\Appendix{Log-periodic functions}

The log-periodic function $c(l)$ defined in Eq.~(\ref{fl}) is such that
$c(b^nl)\!=\!c(l)$ where $b$ is the discrete dilatation factor of the
aperiodic sequence. It can be generally written as a Fourier
expansion,
\begin{eqnarray}
c(l)&=&c_0+\sum_{k=1}^\infty c_k\, \cos[\theta_k(l)]\,,\nonumber\\
\theta_k(l)&=&2\pi k\,\frac{\ln l}{\ln b}+\varphi_k\,,
\label{fecl} 
\end{eqnarray}
so that, in~(\ref{ckj}):
\begin{equation}
c_\kappa(j)=c_0+\sum_{k=1}^\infty c_k
\frac{\sum_{l=1}^jl^{\omega-\kappa-1}\cos\theta_k(l)}{\sum_{l=1}^jl^{\omega-\kappa-1}}\,.
\label{feckj}
\end{equation}

We are interested in the behaviour of $c_\kappa(j)$ at large $j$.

$\bullet$
When $\kappa\!<\!\omega$, let us write:
\begin{equation}
c_\kappa(j)=c_0+\sum_{k=1}^\infty c_k \Re[I_k(j)]\,,
\label{ckj2}
\end{equation}
where:
\begin{eqnarray}
I_k(j)&=&{\rm e}^{{\rm i}\varphi_k}
\frac{\sum_{l=1}^jl^{\omega-\kappa-1+{\rm i}2\pi k/\ln b}}
{\sum_{l=1}^jl^{\omega-\kappa-1}}\nonumber\\
&\simeq&{\rm e}^{{\rm i}\varphi_k}
\frac{\int_0^j{\rm d}l\, l^{\omega-\kappa-1+{\rm i}2\pi k/\ln b}}
{\int_0^j{\rm d}l\, l^{\omega-\kappa-1}}\,,
\label{ikj}
\end{eqnarray}
so that:
\begin{eqnarray}
\Re[I_k(j)]&\simeq&\frac{\cos[\theta_k(j)]+\alpha_k\,\sin[\theta_k(j)]}
{1+\alpha_k^2}\,,\nonumber\\
\alpha_k&=&\frac{2\pi k}{(\omega-\kappa)\ln b}\,.
\label{rikj}
\end{eqnarray}
It follows that $c_\kappa(j)$ oscillates log-periodically around
$c_0$. The Fourier coefficients $c_k$ are divided by $k$ when $k$
is large or when $\kappa$ is close to $\omega$. The effective constant
$\overline{c}$ in~(\ref{lnpj2}) can be taken as the value of
$c_\kappa(j)$ which gives the main contribution to the function in
which it enters.  

$\bullet$
When $\kappa\!=\!\omega$, replacing the sums over $l$ in~(\ref{feckj}) by
integrals, the change of variable $u\!=\!\ln l$ leads to:
\begin{equation}
c_\kappa(j)\simeq c_0+\sum_{k=1}^\infty\frac{c_k}{\ln j}
\int_0^{\ln j}{\rm d}u\,\cos\left(2\pi k\frac{u}{\ln b}
+\varphi_k\right)\,.
\label{ckj1}
\end{equation}
Thus $\lim_{j\to\infty}c_\kappa(j)\!=\!c_0$ and $\overline{c}$
is the constant term in the Fourier expansion of $c(l)$. 

$\bullet$
When $\kappa\!>\!\omega$, according
to~(\ref{ikj}), the asymptotic expression of $c_\kappa(j)$ can
be written in terms of $\zeta$ functions with:
\begin{equation}
I_k(\infty)={\rm e}^{{\rm i}\varphi_k}\frac{\zeta(1+\kappa-\omega 
-{\rm i}2\pi k/\ln b)}{\zeta(1+\kappa-\omega)}\,.
\label{ikj1}
\end{equation}
Thus, in this case too, $c_\kappa(j)$ tends to a well-defined limiting
value giving $\overline{c}$.

\Appendix{Amplitudes and exponents}

We study successively the temperature-dependence of the surface
magnetization near the critical point as well as its size-dependence at
criticality. 

We first consider the case of relevant perturbations
where the leading contribution to $\ln P_j$ is some positive power 
of $j$. The sum $S\!=\!m_{\rm s}^{-2}$ in Eq.~(\ref{pj}) can be then
replaced by an integral of the form~\cite{peschel84,igloi94} 
\begin{equation}
S(t)\simeq\int_0^\infty{\rm d}j\,\lambda^{-2j}P_j^{-2}
\simeq\int_0^\infty{\rm d}j\,\exp(-tj+Bj^\tau)\,, 
\label{st1}
\end{equation}
where we used the definition of $t$ given in~(\ref{t}).
For the finite-size behaviour at the critical point, $t\!=\!0$, the sum
is cut off at L so that:
\begin{equation}
S_{\rm c}(L)\simeq\int_0^L{\rm d}j\,P_j^{-2}
\simeq\int_0^L{\rm d}j\,\exp(Bj^\tau)\,,
\label{sl1}
\end{equation}

When $B\!>\!0$, the integral can be evaluated
using Laplace's method when $0\!<\!\tau\!<\!1$ and, up to a power law
prefactor, one obtains: 
\begin{equation}
m_{\rm s}(t)\sim\exp\left[-\frac{1-\tau}{2\tau}(\tau B)^{1/(1-\tau)}\, 
t^{-\tau/(1-\tau)}\right]\,.
\label{mst4}
\end{equation}
The main contribution to $S_{\rm c}(L)$ comes from the vicinity of the
upper limit. Expanding the argument of the exponential around $L$
leads to:
\begin{equation}
m_{\rm s,c}(L)\sim\exp\left[-\frac{1}{2}B\, L^\tau\right]\,.
\label{msl4}
\end{equation}

When $B\!<\!0$, expanding $\exp(-tj)$ in~(\ref{st1}) and integrating
term by term gives:
\begin{equation}
m_{\rm s}(t)\simeq\left[\frac{\tau\vert B\vert^{1/\tau}}{\Gamma(1/\tau)}
\right]^{1/2}\left[1+\frac{\Gamma(2/\tau)}{2\Gamma(1/\tau)
\vert B\vert^{1/\tau}}\, t+\cdots\right]\,.
\label{mst5}
\end{equation}

Let us now look for the values of $B$ and $\tau$ when
$\kappa$ varies.

$\bullet$
When $1-\omega\!<\!\kappa\!<\!\omega$, the term which is linear in $a$
dominates in~(\ref{lnpj2}), so that:
\begin{equation}
P_j^{-2}\simeq\exp\left[\frac{4\overline{c}a}{\omega-\kappa}
j^{\omega-\kappa}\right]\,,
\label{b1}
\end{equation}
which leads to the expressions given in Eqs.~(\ref{mst1})
and~(\ref{msl1}).

$\bullet$
When $1-\omega\!=\!\kappa\!<\!\omega$, both terms in~(\ref{lnpj2}) are of
the same order and
\begin{equation}
P_j^{-2}\simeq\exp\left[\frac{4\overline{c}a+a^2}{1-2\kappa}
j^{1-2\kappa}\right]\,,
\label{b2}
\end{equation}
leading to~(\ref{mst2}) and~(\ref{msl2}).

$\bullet$
When $\kappa\!<\!{\rm min}(1-\omega,1/2)$, the term in $a^2$
governs the behaviour of~(\ref{lnpj2}), so that
\begin{equation}
P_j^{-2}\simeq\exp\left[\frac{a^2}{1-2\kappa}
j^{1-2\kappa}\right]\,,
\label{b3}
\end{equation}
from which Eqs.~(\ref{mst3}) and~(\ref{msl3}) follow. 

Next we consider the case of marginal perturbations where 
$P_j$ behaves as $j^{-B/2}$. When $B\!>\!-1$, $S$ in Eq.~(\ref{pj}) 
can be rewritten as~\cite{peschel84}    
\begin{equation}
S(t)\simeq\int_0^\infty{\rm d}j\,\lambda^{-2j}P_j^{-2}
\simeq\int_0^\infty{\rm d}j\, j^{B}{\rm e}^{-tj}\,, 
\label{st2}
\end{equation}
or, at the critical point,
\begin{equation}
S_{\rm c}(L)\simeq\int_0^L{\rm d}j\,P_j^{-2}
\simeq\int_0^L{\rm d}j\, j^{B}\,.
\label{sl2}
\end{equation}
For the $t$-dependence, one obtains:
\begin{eqnarray}
m_{\rm s}(t)&\simeq&\left[t^{-1-B}\int_0^\infty{\rm d}u\, 
u^B{\rm e}^{-u} \right]^{-1/2}\nonumber\\
&=&[\Gamma(B+1)]^{-1/2}\, t^{(1+B)/2}\,,
\label{mst6}
\end{eqnarray}
whereas:
\begin{equation}
m_{\rm s,c}(L)\simeq (1+B)^{1/2}L^{-(1+B)/2}
\label{msl6}
\end{equation}
When $B\!<\!-1$, the integrals in~(\ref{st2}) and~(\ref{sl2})
diverge at their lower limits. The main contribution, coming
from small values of $j$, must be treated more
carefully. For this purpose let us split $S$ into two parts
as:   
\begin{equation}
S(t)=1+\sum_{j=1}^\infty j^{-\vert B\vert}
+\sum_{j=1}^\infty({\rm e}^{-tj}-1)j^{-\vert B\vert}\,.
\label{st3}
\end{equation}
The first sum gives $\zeta(\vert B\vert)$ whereas
the second sum can be transformed using the 
Euler-MacLaurin sum\-mation formula:
\begin{equation}
\sum_{j=1}^\infty f(j,t)
=\int_1^{t^{-1}}\!\!{\rm d}j\, f(j,t)
+\int_{t^{-1}}^\infty\!\!{\rm d}j\, f(j,t)+O(t)\,.
\label{eml}
\end{equation}
The change of variable $u\!=\!tj$ leads to:
\begin{eqnarray}
S(t)&=&m_{\rm s,c}^{-2}+t^{\vert B\vert-1}\Big[
\int_t^1\!\!{\rm d}u\,({\rm e}^{-u}-1)u^{-\vert B\vert}
+\nonumber\\
&&\qquad\qquad+\int_1^\infty\!\!{\rm d}u\,
({\rm e}^{-u}-1)u^{-\vert B\vert}\Big]\,,
\label{cv}
\end{eqnarray}
where $m_{\rm s,c}\!=\![1+\zeta(\vert B\vert)]^{-1/2}$.
The $t$-dependence of the first integral is obtained through an 
expansion of the exponential and the second is a constant. Collecting the 
different terms gives
\begin{equation}
S(t)=m_{\rm s,c}^{-2}+{\rm const}\, t^{\vert B\vert-1}+O(t)
\label{cv1}
\end{equation}
and
\begin{equation}
m_{\rm s}(t)=m_{\rm s,c}+{\rm const}\, t^{\vert B\vert-1}+O(t)\,.
\label{mst7}
\end{equation}
For the size-dependence at criticality, one may write:
\begin{eqnarray}
S_{\rm c}(L)&=&1+\sum_{j=1}^Lj^{-\vert B\vert}\nonumber\\
&=&1+\sum_{j=1}^\infty
j^{-\vert B\vert}-\int_L^\infty{\rm d}j\, j^{-\vert B\vert}
+O(L^{-\vert B\vert})\nonumber\\
&=&m_{\rm s,c}^{-2}+\frac{L^{1-\vert B\vert}}{1-\vert B\vert}
+O(L^{-\vert B\vert})\,,
\label{sl3}
\end{eqnarray}
so that:
\begin{equation}
m_{\rm s,c}(L)=m_{\rm s,c}+
\frac{m_{\rm s,c}^3}{2(\vert B\vert-1)}\,L^{-\vert B\vert+1}\,. 
\label{msl7}
\end{equation}

Finally, we identify the expression of the exponent $B$
for $\kappa$-values where a marginal
behaviour is obtained. 

$\bullet$
When $\kappa\!=\!\omega\!>\!1/2$, the linear term
in~(\ref{lnpj2}) is the dominant one and yields:
\begin{equation}
P_j^{-2}\simeq j^{4\overline{c}a}\,,
\label{b4}
\end{equation}
from which the exponents in~(\ref{betas1})
and~(\ref{betas2}) follow.  

$\bullet$
When $\kappa\!=\!\omega\!=\!1/2$, the linear and quadratic terms 
in Eq.~(\ref{lnpj2}) contribute, so that
\begin{equation}
P_j^{-2}\simeq j^{4\overline{c}a+a^2}\,,
\label{b5}
\end{equation}
leading to~(\ref{betas3}) and~(\ref{betas4}).

$\bullet$
When $\kappa\!=\!1/2$ and $\omega\!<\!1/2$, the leading
contribution in~(\ref{lnpj2}) is the quadratic one,
\begin{equation}
P_j^{-2}\simeq j^{a^2}\,,
\label{b6}
\end{equation}
which gives the exponent in Eq.~(\ref{betas5}).


\begin{thebibliography}{}
\bibitem{gap} One may notice that for a symmetric system, where both 
surfaces are perturbed, the first gap scales as 
$L^{-1/2+\beta_{\rm s}(a)}$. See for example Ref.~\cite{karevski95}.
\end{thebibliography}
\end{document}